# Vulnerability-Weighted Routing of Timing-Critical Nets for Configuration-Upset–Resilient SRAM-Based FPGAs

Mostafa Darvishi, *Senior Member, IEEE*

***Abstract*— Conventional FPGA routing optimizes timing, congestion, and routability but does not distinguish routes with similar nominal performance and substantially different susceptibility to configuration-induced delay degradation. This paper presents a vulnerability-weighted routing methodology for SRAM-based field-programmable gate arrays (FPGAs) that incorporates predicted routing-fault severity directly into the routing objective. A continuous vulnerability cost relates delay perturbations caused by electrically attachable dormant routing resources to the available downstream timing slack, while a complementary configuration-concentration term discourages excessive localization of vulnerable resources. To limit implementation disruption, only the highest-risk nets are selectively ripped up and rerouted while unaffected routes remain fixed. The method is implemented on a Zynq UltraScale+ XCZU7EV using a Vivado/RapidWright-based flow and evaluated across four routed benchmarks against commercial timing-driven routing, vulnerability-agnostic rerouting, and binary vulnerable-resource avoidance. Controlled configuration-equivalent perturbations provide hardware-level validation. The proposed method reduces aggregate configuration-induced timing vulnerability by 41.7% with approximately 1.0% nominal timing degradation and captures 85.8% of the vulnerability reduction obtained at the expanded routing budget by rerouting only the highest-risk 5% of eligible nets. The results demonstrate that continuous vulnerability information can improve configuration-upset resilience with limited impact on nominal routing quality.**

## I. INTRODUCTION

Routing is a dominant stage of FPGA physical implementation because the selected programmable interconnection resources determine routability, congestion, and a substantial portion of the final path delay. Classical FPGA routers such as *PathFinder* [1] formulate routing as a negotiated optimization between performance and congestion, while *VPR* [2] established a widely used framework for architecture-aware placement and routing research. Modern commercial-device flows retain the same fundamental requirement of obtaining legal routes while satisfying timing objectives, and open physical-design frameworks such as *RapidWright* [3] and *RWRoute* [4] now permit customized timing-driven routing directly on commercial UltraScale+ devices.

For SRAM-based FPGAs, however, nominal routing quality is not the only relevant design objective. Configuration memory controls the programmable interconnection fabric, and an upset affecting routing control may modify the electrical environment of an otherwise functional net [5]. Circuit-level modeling [6] and radiation experiments [7] have shown that unintentionally attached programmable interconnection points (PIPs) and dormant wire segments can introduce additional loading and increase combinational delay. The resulting perturbation may remain logically silent while consuming path slack, making a route vulnerable even when its fault-free timing satisfies the design constraint [8].

Reliability-aware CAD techniques [9] have previously modified FPGA placement and routing to mitigate configuration-upset effects, and SEU-aware routing has been used specifically to reduce upset-induced bridging faults [8]. These methods demonstrate that physical routing decisions can materially influence FPGA reliability. Nevertheless, their objectives do not explicitly distinguish two legal routes having comparable nominal delay when the routes are exposed to different continuous distributions of configuration-induced delay severity. In particular, a route containing many electrically significant dormant attachments near low-slack segments should not necessarily receive the same routing preference as an equally fast route whose neighboring programmable resources produce substantially smaller timing perturbations.

This distinction is illustrated in Fig. *1*(a). *Route A* and *Route B* may satisfy essentially the same nominal timing requirement, yet *Route A* traverses resources associated with more numerous or more severe dormant attachments. Conventional timing/congestion optimization has no direct representation of this difference. The proposed methodology therefore augments the routing objective with a continuous timing-vulnerability term and a configuration-concentration term, as summarized in Fig. *1*(b). Rather than globally rerouting the complete implementation, the flow selectively identifies high-risk nets, removes their existing routes, and searches for alternatives that reduce configuration-induced slack-exhaustion risk while retaining acceptable timing and congestion.

The resulting edge cost jointly considers nominal delay, congestion, predicted routing-fault vulnerability, and

Mostafa Darvishi (S'06–M'18–SM'25) is with Electrical Engineering Department of École de technologie supérieure (ÉTS), Montreal, Canada. (e-mail: darvishi@ieee.org).

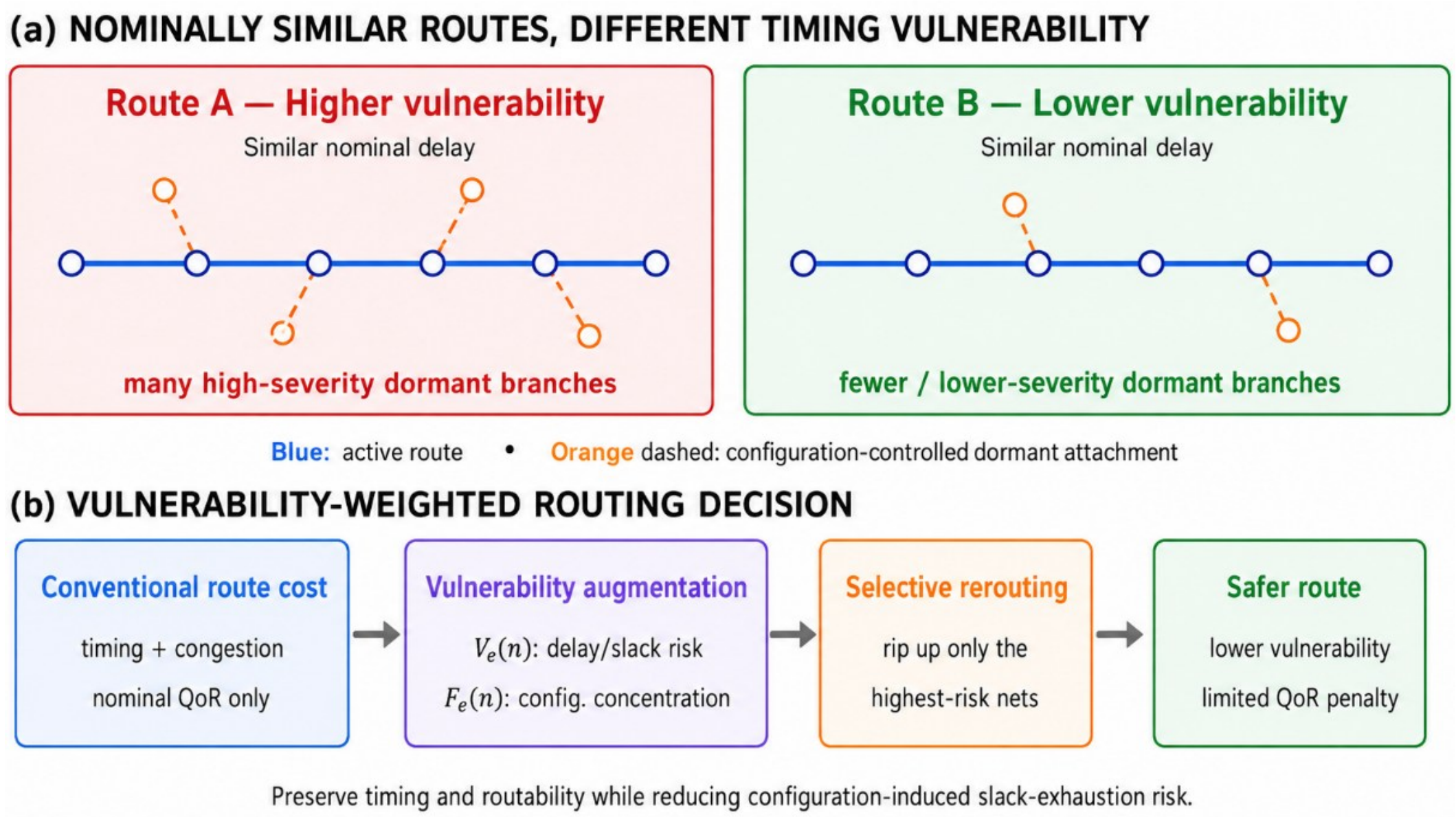

Fig. 1. Motivation for vulnerability-weighted FPGA routing. **(a)** Two routes with comparable nominal timing can exhibit different susceptibility to configuration-induced delay because of the number and severity of electrically attachable dormant routing resources. **(b)** The proposed flow augments conventional timing/congestion cost with routing-vulnerability and configuration-concentration terms and selectively reroutes high-risk nets toward safer alternatives.

concentration of exposed resources within common configuration regions. Vulnerability is derived from the delay consequence of candidate dormant-resource attachments relative to the available timing slack, rather than from binary classification of a resource as simply *critical* or *noncritical*. This formulation enables routing decisions to distinguish between electrically benign and timing-threatening configuration perturbations while preserving the timing-driven behavior required for practical FPGA implementation. *RapidWright* provides access to physical routing resources in commercial devices [3], while *RWRoute* [4] demonstrates that a customizable timing-driven routing framework can operate on UltraScale+ architectures, providing a practical foundation for the proposed selective-routing flow.

The method is evaluated on four structurally different benchmark designs implemented on a Zynq UltraScale+ XCZU7EV FPGA. Commercial timing-driven routing, vulnerability-agnostic custom routing, and binary vulnerable-resource avoidance are used as baselines. In addition to post-route timing, congestion, wirelength, and vulnerability analysis, controlled function-preserving routing perturbations are applied to a stratified subset of baseline and rerouted paths to determine whether routes assigned lower vulnerability scores also exhibit lower measured configuration-induced timing degradation.

The main contributions of this work are: **1)** a continuous routing-resource vulnerability metric that relates candidate configuration-induced branch attachments to predicted delay and available path slack; **2)** a vulnerability-weighted routing cost that jointly considers nominal timing, congestion, timing-fault severity, and configuration-resource concentration; **3)** a selective rip-up-and-reroute strategy that concentrates reliability optimization on the highest-risk nets to limit degradation of nominal quality of results; and **4)** a multi-benchmark implementation and hardware-validation methodology comparing vulnerability reduction against timing, routing, and computational overhead.

The remainder of this paper is organized as follows. Section II reviews FPGA routing and reliability-aware physical-design methods. Section III defines the configuration-induced routing-vulnerability model. Section IV presents the proposed vulnerability-weighted routing algorithm. Section V describes the experimental platform, benchmarks, baselines, and controlled perturbation methodology. Section VI presents the experimental results, Section VII discusses practical implications and limitations, and Section VIII concludes the paper.

## II. BACKGROUND AND RELATED WORK

### A. Timing-Driven FPGA Routing

FPGA routing is commonly represented using a routing-resource graph in which programmable wires and switches form candidate paths between source and sink terminals [7]. *PathFinder* [1] established the negotiated-congestion formulation in which routing cost evolves according to resource demand while performance-sensitive connections are weighted according to timing criticality. *VPR* [2] subsequently provided a widely used architecture and CAD framework for evaluating timing-driven placement and routing techniques. These approaches established the fundamental tradeoff among routability, congestion, and nominal path delay, but their routing objectives contain no explicit representation of configuration-upset susceptibility.

For commercial devices, *RapidWright* [3] exposes physical routing resources and permits manipulation of placed-and-routed designs. *RWRoute* [4] extends this capability with a timing-driven router targeting UltraScale+ devices and incorporates a device-specific delay model suitable for timing-aware route selection. These capabilities provide the physical-design infrastructure required for the present work; however, the routing objective itself remains principally concerned with timing and congestion rather than configuration-induced timing vulnerability.

### *B. Reliability-Aware FPGA Physical Design*

Several prior studies have demonstrated that FPGA reliability can be improved through physical-design decisions. Sterpone and Violante [9] introduced a reliability-oriented place-and-route methodology for SRAM-based FPGAs that restructures placement and routing to reduce failures associated with configuration-memory upsets. Golshan and Bozorgzadeh [8] subsequently proposed an SEU-aware router aimed specifically at reducing routing bridging faults by decreasing the number of configuration care bits capable of producing such faults. Their reported experiments showed that routing choices can reduce upset susceptibility while retaining comparable nominal delay.

Abdul-Aziz and Tahoori [10] incorporated soft-error reliability into modified *VPR* placement and routing algorithms while preserving conventional area and delay objectives. Their formulation focuses on reducing the number of sensitive configuration bits. Huang *et al.* [11] extended reliability-oriented physical design by jointly considering fault-occurrence probability and error-propagation probability, thereby recognizing that an upset is relevant only if its effect can propagate to system outputs. These approaches move beyond purely geometric avoidance, but their reliability quantities are primarily associated with functional soft-error manifestation rather than the magnitude of a configuration-induced timing-delay perturbation relative to the remaining path slack.

Complementary work has focused on vulnerability analysis rather than route optimization. Sari *et al.* [12] developed an open-source framework for estimating soft-error susceptibility in Xilinx FPGA implementations and for evaluating reliability-aware CAD techniques. Such analysis frameworks are important for identifying sensitive configuration resources, but vulnerability estimation alone does not determine how a timing-driven router should trade additional nominal routing cost against the continuous severity of alternative routing-delay faults.

### *C. Position of the Proposed Method*

Table I summarizes the distinction between representative prior approaches and the present work. Earlier reliability-aware routing methods primarily optimize sensitive-bit count, bridging-fault susceptibility, fault occurrence, or functional error propagation [8], [9], [10], [11]. In contrast, the proposed method assigns each routing alternative a **continuous timing-vulnerability cost** derived from the predicted delay increment of configuration-induced dormant-resource attachments relative to available timing slack. It additionally penalizes excessive concentration of vulnerable routing resources within common configuration regions and applies this cost selectively to the highest-risk nets rather than rerouting the complete implementation.

This distinction is important because two routes with similar nominal delay, sensitive-bit count, or functional connectivity can still differ substantially in the timing consequence of a configuration perturbation. The proposed objective therefore complements conventional timing and congestion costs rather than replacing them. Moreover, by building the methodology on *RapidWright*/*RWRoute*-compatible UltraScale+ routing resources [3], [4], the evaluation is performed on a commercial FPGA architecture instead of an abstract island-style fabric. The contribution is thus positioned at the intersection of **timing-driven commercial-device routing** and **configuration-aware reliability optimization**, with delay severity and remaining slack forming the principal reliability metric.

## III. CONFIGURATION-INDUCED ROUTING VULNERABILITY MODEL

### *A. Routed-Net and Perturbation Model*

The post-route programmable interconnection fabric is represented by a directed graph

$$\mathcal{G} = (\mathcal{V}, \mathcal{E}), \tag{1}$$

where $\mathcal{V}$ denotes routing nodes and $\mathcal{E}$ denotes programmable routing connections. For a routed net $n$, let $\mathcal{R}_n \subseteq \mathcal{E}$ denote the set of routing resources forming its source-to-sink routing tree. This representation is consistent with the physical routing-resource abstraction exposed by *RapidWright*, which provides explicit wire, node, and PIP objects for commercial Xilinx/AMD devices [3]. For every active routing edge $e \in \mathcal{R}_n$, the set

$$\mathcal{B}_e = \{b_1, b_2, \dots, b_{N_e}\} \tag{2}$$

contains configuration-induced delay candidates electrically adjacent to that resource. Candidate $b$ is described by

$$b = (a_b, p_b, H_b, \phi_b, \widehat{\Delta d}_b, \sigma_b) \tag{3}$$

where $a_b$ is the attachment point on the active route, $p_b$ is the inactive PIP whose unintended activation creates the perturbation, $H_b$ is the attached dormant routing component, and $\phi_b$ identifies its verified configuration-frame or configuration-region association. The quantities $\widehat{\Delta d}_b$ and $\sigma_b$ respectively denote the predicted delay increment and its calibrated uncertainty. Only perturbations that preserve intended Boolean connectivity while modifying the electrical loading of the routed net are retained. Candidates producing opens, logical shorts, or additional active drivers are excluded because they correspond to functional faults rather than the delay-only mechanism considered here [5], [6].

UltraScale/UltraScale+ programmable logic is controlled through frame-organized configuration memory [13].

Table I. Comparison of FPGA routing and reliability-aware CAD approaches

| Approach | Main Objective | CU-Aware | Delay/Slack Severity | Config. Concentration | UltraScale+ | Selective Rerouting |
|---|---|---|---|---|---|---|
| **PathFinder &VPR [1], [2]** | Timing, congestion, routability | — | — | — | — | — |
| **Sterpone & Violante [9]** | Reliability-oriented place/route | ✓ | — | — | — | — |
| **Golshan & Bozorgzadeh [8]** | SEU-induced bridging-fault reduction | ✓ | — | — | — | — |
| **Abdul-Aziz & Tahoori [10]** | Sensitive-bit reduction + area/delay | ✓ | — | — | — | — |
| **Huang *et al.* [11]** | Fault occurrence + error propagation | ✓ | — | — | — | — |
| **Sari *et al.* [12]** | Soft-error vulnerability analysis | ✓ | — | — | — | — |
| **RWRoute [4]** | Timing-driven commercial-device routing | — | — | — | ✓ | ◐ |
| **Proposed Method** | **Timing + congestion + continuous vulnerability** | ✓ | ✓ | ✓ | ✓ | ✓ |

**Abbreviations**: **CU** = configuration-upset; ✓: explicitly supported; ◐: available indirectly but not a primary optimization objective; —: not a primary feature of the referenced method.

Accordingly, $\phi_b$ is assigned only where the configuration association can be established by the extraction and mapping procedure described later in Section V; an unmapped candidate remains part of the delay-vulnerability calculation but is omitted from the configuration-concentration term. This prevents an assumed bit-to-resource correspondence from being introduced solely for optimization.

### *B. Slack-Exhaustion Severity*

A routing perturbation is relevant to timing reliability only when its additional delay is significant relative to the slack available downstream of its attachment point. Let $\mathcal{D}_b$ denote the set of timing endpoints whose paths contain the affected routing segment. The effective slack associated with candidate *b* is therefore defined conservatively as

$$S_b = \min_{s\in\mathcal{D}_b} S_s \tag{4}$$

where $S_s$ is the post-route setup slack of endpoint *s*. This endpoint-specific definition is particularly important for multi-sink routing trees, because a perturbation introduced on one branch affects only the downstream sinks traversing that branch; using a net-wide slack value could therefore overestimate or underestimate its timing consequence. The severity of candidate *b* is then expressed as the probability that its delay increment exceeds this available margin,

$$w_b = P(\Delta d_b > S_b) = 1 - F_{\Delta d_b}(S_b), \qquad 0 \le w_b \le 1 \tag{5}$$

where $F_{\Delta d_b}(\cdot)$ is the calibrated cumulative distribution of the predicted delay perturbation. When the prediction residual is adequately represented by a Gaussian distribution,

$$1 - \Phi\left(\frac{S_b - \widehat{\Delta d}_b}{\sigma_b}\right) \tag{6}$$

with $\Phi(\cdot)$ denoting the standard-normal cumulative distribution function. Thus, two electrically possible perturbations can receive substantially different severity even when they attach to routing resources of the same type: a small delay increase on a low-slack path can be more consequential than a larger perturbation on a path having substantial timing margin. Importantly, $w_b$ is a **conditional timing-severity quantity**, not the probability that radiation produces configuration upset *b*. It therefore ranks the timing consequence *given* the considered configuration perturbation and does not require or imply a device-level upset cross section. Radiation-event probability can subsequently be incorporated as an external exposure weight when such application-specific data are available.

### *C. Edge and Route Vulnerability*

The continuous vulnerability associated with routing edge *e* for net *n* is defined as

$$\sum_{b\in\mathcal{B}_e} w_b \tag{7}$$

and the aggregate configuration-induced timing vulnerability of route $\mathcal{R}_n$ is

$$\sum_{e\in\mathcal{R}_n} V_e(n) \tag{8}$$

Consequently, an edge is penalized not merely because configurable resources exist nearby, but according to both the number of admissible delay-producing attachments and their predicted ability to consume the downstream timing margin. Equation (8) is used as a comparative routing metric rather than as an absolute system-failure probability. A second quantity captures how the vulnerability of a route is distributed across configuration regions. For configuration region *f*, let us define its vulnerability load as

$$Q_{n,f} = \sum_{e \in \mathcal{R}_n} \sum_{\substack{b \in \mathcal{B}_e \\ \phi_b = f}} w_b. \tag{9}$$

Here, $Q_{n,f}$ represents the total timing-severity contribution of all mapped routing-fault candidates associated with configuration region $f$ along route $\mathcal{R}_n$. In other words, $Q_{n,f}$ defines **how much vulnerability lies in region *f*.** Let

$$\mathcal{F}_n = \{f : Q_{n,f} > 0\}$$

denote the set of mapped configuration regions that contribute nonzero vulnerability to route $\mathcal{R}_n$. For $\mathcal{F}_n \neq \emptyset$, the configuration-concentration index is then defined as

$$F(\mathcal{R}_n) = \sum_{f \in \mathcal{F}_n} \left( \frac{Q_{n,f}}{\sum_{g \in \mathcal{F}_n} Q_{n,g}} \right)^2. \tag{10}$$

The index satisfies

$$\frac{1}{|\mathcal{F}_n|} \leq F(\mathcal{R}_n) \leq 1. \tag{11}$$

The lower bound occurs when the mapped vulnerability is uniformly distributed among all configuration regions in $\mathcal{F}_n$, whereas $F(\mathcal{R}_n) = 1$ when the entire mapped vulnerability is concentrated in a single configuration region. Therefore, larger values of $F(\mathcal{R}_n)$ indicate stronger concentration of vulnerable routing resources within a smaller portion of configuration space. If $\mathcal{F}_n = \emptyset$, meaning that none of the vulnerability-contributing candidates along $\mathcal{R}_n$ has a verified configuration-region association, the concentration term is defined as

$$F(\mathcal{R}_n) = 0,$$

and is excluded from the configuration-concentration penalty. Such candidates still contribute normally to the aggregate route vulnerability $V(\mathcal{R}_n)$ through (7) and (8).

The configuration-concentration term is intentionally treated as a secondary physical-design penalty. It does not imply that a single configuration-bit upset simultaneously activates all candidates mapped to the same region; rather, it discourages routing solutions whose cumulative vulnerable resources are excessively concentrated within a limited number of common configuration regions. AMD documentation confirms the frame-organized configuration structure relevant to UltraScale-class devices and identifies the Zynq UltraScale+ technical reference as the device-specific configuration reference [13].

Equations (7)–(10) therefore provide two complementary route properties: $V(\mathcal{R}_n)$ measures the aggregate severity of configuration-induced timing perturbations, whereas $F(\mathcal{R}_n)$ measures their configuration-space concentration. Section IV incorporates both quantities into the timing- and congestion-aware selective-routing objective.

## IV. PROPOSED VULNERABILITY-WEIGHTED ROUTING

### *A. Selective High-Risk Net Identification*

Let $\mathcal{N}_{\text{elig}}$ denote the set of signal nets that are legally available for rerouting after excluding clock networks, dedicated interconnects, static nets, and routes intentionally locked by the implementation flow. For each eligible net *n*, the baseline route vulnerability is

$$V_n = V(\mathcal{R}_n) \tag{12}$$

where $V(\mathcal{R}_n)$ is defined by (8). Because the candidate severities $w_b$ already incorporate the downstream timing slack, $V_n$ jointly reflects the number and timing consequence of configuration-induced routing perturbations. A second slack-dependent factor is therefore not applied during net ranking. For a prescribed rerouting fraction *p*, the selected population is

$$\text{Top}_K\{V_n \in \mathcal{N}_{\text{elig}}\}, \qquad K = \lceil p|\mathcal{N}_{\text{elig}}| \rceil. \tag{13}$$

The remaining nets retain their original placement and routing. Varying *p* permits direct exploration of the tradeoff between implementation disruption and achievable vulnerability reduction; Section VI evaluates this tradeoff over multiple rerouting fractions.

### *B. Vulnerability-Augmented Routing Cost*

For a candidate route $\mathcal{R}_n$, the proposed routing objective is

$$\alpha\overline{D}(\mathcal{R}_n) + \beta\overline{G}(\mathcal{R}_n) + \gamma\overline{V}(\mathcal{R}_n) + \delta\overline{F}(\mathcal{R}_n) \tag{14}$$

where *D* denotes timing-driven route cost, *G* denotes negotiated congestion cost, *V* is the configuration-induced timing vulnerability defined in (8), and *F* is the configuration-concentration index of (10). The overbar denotes normalization to a common dimensionless range, and

$$\alpha, \beta, \gamma, \delta \geq 0, \qquad \alpha + \beta + \gamma + \delta = 1. \tag{15}$$

Normalization parameters are obtained from the legal candidate-resource population and held fixed for all routing alternatives of a given experimental configuration. Consequently, the weighting coefficients determine the intended design tradeoff rather than compensating for incompatible numerical scales. During routing iteration *k*, expansion through candidate edge *e* uses the incremental cost

$$\alpha\overline{D}_e^{(k)}(n) + \beta\overline{G}_e^{(k)} + \gamma\overline{V}_e(n) + \delta\overline{\Delta F}_e^{(k)}(n) \tag{16}$$

where $\overline{D}e^{(k)}(n)$ retains the timing-driven criticality and delay contribution of the underlying router, and $\overline{G}e^{(k)}$ retains its negotiated resource-use penalty. Thus, the proposed method augments rather than replaces conventional timing/congestion routing [1], [8]. The edge vulnerability follows directly from (7). Because configuration concentration is a route-level quantity, its incremental contribution is evaluated as

$$F\left(\mathcal{R}_n^{(k)}\right) \tag{17}$$

where $\mathcal{R}_n^{(k)}$ is the partial route constructed at the current search state. The search penalty uses the normalized positive increase

$$\max\left(0, \Delta F_e^{(k)}\right) \tag{18}$$

thereby avoiding negative edge costs while discouraging moves that further concentrate vulnerability within already dominant configuration regions. The complete candidate route is subsequently evaluated using the full $F(\mathcal{R}_n)$ term in (13), so alternatives that ultimately distribute vulnerability more favorably can still be preferred. Candidates lacking a verified configuration association, contribute to $V$ but not to $F$, as defined in Section III.

### C. Selective Rip-Up-and-Reroute Procedure

The proposed flow starts from a timing-closed routed design rather than routing the complete netlist from scratch. The baseline physical design is imported into the *RapidWright*/*RWRoute* environment, and the quantities $V_n$ and $F(\mathcal{R}_n)$ are evaluated for all eligible nets. Nets in $\mathcal{N}_p$ are then processed in descending baseline vulnerability, while routes outside the selected population are preserved.

For each selected net, the original route is temporarily removed, and candidate connections are reconstructed using (15). Negotiated congestion costs continue to evolve across routing iterations, preventing vulnerability reduction from being obtained through persistent resource overuse. If no legal alternative satisfying the timing and routing constraints is found, the original route is restored. The procedure can therefore leave an already favorable route unchanged rather than forcing a reliability modification that degrades implementation quality. Algorithm 1 summarizes the entire process. This procedure preserves the placement and the routes of nonselected nets, concentrating computational effort and physical-design disruption on the portion of the implementation that contributes most strongly to configuration-induced timing vulnerability.

**Algorithm 1** Selective Vulnerability-Weighted Rerouting

**1**: **Input:** routed design, $(B)$, $(p)$, and $(\alpha, \beta, \gamma, \delta)$.
**2**: Extract $(\mathcal{N}_{\text{elig}})$ and compute $(V_n)$ for each eligible net.
**3**: Form $(\mathcal{N}_p)$ using (12); preserve all unselected routes.
**4**: Sort $(\mathcal{N}_p)$ in descending $(V_n)$.
**5**: **For** each $(n \in \mathcal{N}_p)$:
  a) save and rip up $(\mathcal{R}_n)$;
  b) search candidate routes using the cost in (15);
  c) update negotiated congestion costs;
  d) retain the best legal candidate satisfying the acceptance constraints; otherwise restore $(\mathcal{R}_n)$.
**6**: Verify routing legality and timing and recompute $(V)$ and $(F)$.
**7**: **Output:** vulnerability-reduced routed design.

### D. Timing Preservation and Route Acceptance

Reliability improvement is accepted only if it does not invalidate the nominal implementation. Let $\mathcal{R}_n^0$ and $\mathcal{R}_n'$ denote the baseline and proposed routes of selected net $n$, respectively. A candidate route must first provide an actual vulnerability benefit,

$$V(\mathcal{R}_n') < V(\mathcal{R}_n^0). \tag{19}$$

At the complete-design level, it must additionally satisfy

$$\text{WNS}' \geq \text{WNS}^0 - \varepsilon_{\text{WNS}}, \qquad N'_{\text{overuse}} = 0 \tag{20}$$

where $\varepsilon_{\text{WNS}}$ is the permitted degradation of worst negative slack and $N_{\text{overuse}}$ is the number of illegally shared routing resources. The placement is held unchanged, and

$$\mathcal{R}_n^0, \qquad \forall n \notin \mathcal{N}_p \tag{21}$$

is verified after routing. Final legality, routed timing, and route preservation are independently checked using the implementation flow described later in Section V.

Equations (13)–(20) therefore establish a bounded reliability optimization: *the router is permitted to exchange a controlled amount of nominal routing quality for lower configuration-induced timing vulnerability, but it cannot obtain that reduction through illegal congestion, uncontrolled timing degradation, or modification of nets outside the selected protection set*.

Fig. 2 summarizes the proposed vulnerability-weighted routing flow. Starting from the routed design, candidate routing perturbations are first enumerated and scored according to their predicted timing consequence, structural exposure, and available slack. These candidate scores are then aggregated at the net level to construct a vulnerability ranking that identifies the routing resources most likely to cause timing failure under configuration-induced perturbations. The routing stage uses this ranking to selectively protect the most critical nets by steering them away from highly exposed regions and by favoring lower-risk routing alternatives, while noncritical nets continue to follow the nominal implementation flow. Finally, the routed solution is validated through timing analysis and in-situ observability measurements, enabling iterative refinement of the vulnerability model and confirming that the proposed framework improves timing robustness without requiring blanket hardening of the entire design.

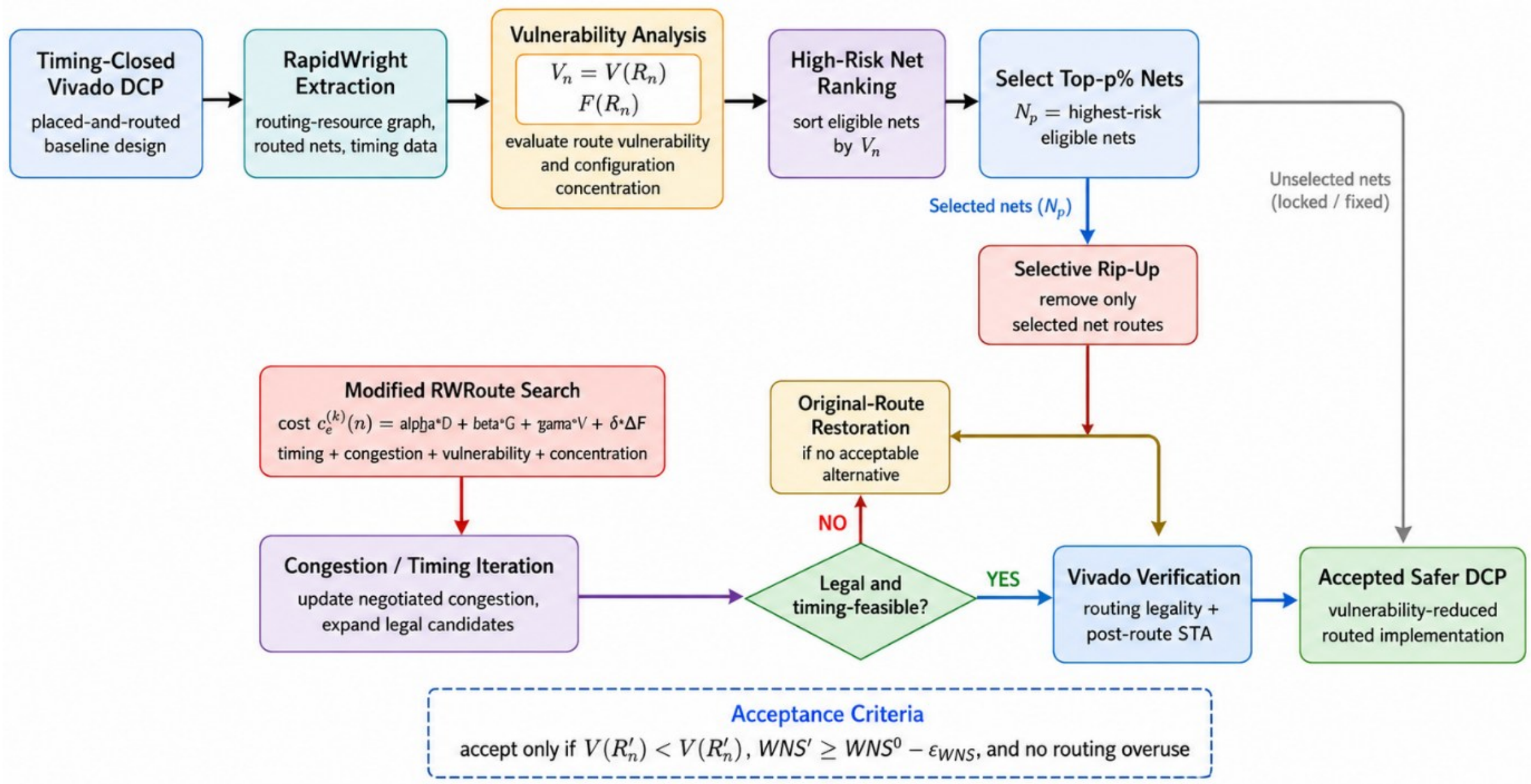

Fig. 2. Proposed vulnerability-weighted routing methodology. Candidate perturbations are scored from timing and structural features, aggregated into net-level vulnerability metrics, and used to guide selective routing decisions for improved robustness against configuration-induced delay degradation.

## V. EXPERIMENTAL METHODOLOGY AND BENCKMARKS

### *A. FPGA Platform and Physical-Design Flow*

Experiments target the AMD ZCU104 evaluation board containing a Zynq UltraScale+ XCZU7EV-2FFVC1156 MPSoC [13]. The reference implementations are synthesized, placed, and routed using Vivado Design Suite 2025.1. Vivado implementation produces the placed-and-routed design checkpoints used as the baseline for the proposed selective-routing flow [14]. Post-route timing is evaluated from the final implementation using the same constraints for all compared routing strategies [15].

Each baseline checkpoint is imported into *RapidWright*, where routed nets, nodes, PIPs, timing information, and neighboring dormant routing resources are extracted. *RapidWright* exposes the commercial-device routing representation, while *RWRoute* provides timing-driven routing and permits the resulting route to be validated again with Vivado [3], [4]. Its partial-routing capability is particularly suitable for the proposed experiment because selected connections can be rerouted while previously routed resources remain preserved.

For every benchmark, five independently implemented baseline instances are generated using distinct implementation seeds. Within each instance, placement is subsequently frozen so that the *commercial baseline*, *vulnerability-agnostic rerouting*, *binary avoidance*, and *proposed vulnerability-weighted rerouting* operate from the same physical placement. After each custom-routing experiment, the checkpoint is returned to Vivado for routing-legality verification and static timing analysis. A result is accepted only if no illegal resource overuse is present and the timing-preservation condition in (19) is satisfied.

### *B. Benchmark Suite*

Four benchmarks are used to exercise distinct routing structures, resource types, fan-out patterns, and timing regimes, as summarized in Fig. 3. **B1** is a five-stage RV32I processor organized into IF, ID, EX, MEM, and WB stages separated by pipeline registers. Its datapath contains a register file, ALU, branch-comparison logic, and data-memory interface, while the instruction path connects the program counter and instruction memory to the processor pipeline. The resulting combination of control-dependent, arithmetic, branch, and memory-access paths provides a heterogeneous routing structure for evaluating vulnerability-aware rerouting. **B2** is an eight-lane 32-bit fixed-point multiply–accumulate (MAC) datapath. Two eight-lane input vectors are registered before entering eight parallel multiplier lanes; their products are combined through a pipelined reduction tree and forwarded to a final accumulation/output stage. This benchmark emphasizes regular arithmetic routing, DSP-adjacent connectivity, and comparatively tight timing constraints.

**B3** is an AXI4-Stream-to-dual-port-BRAM buffering subsystem. The slave-side AXI4-Stream input first passes through width/alignment and packing logic, followed by write-control and address-generation circuitry connected to one port of the dual-port BRAM. An independent read-control and address-generation path accesses the second BRAM port and forwards the retrieved data through an unpack/width-conversion stage to the master-side AXI4-Stream output. Shared address/buffer control logic coordinates the streaming and memory-access functions. **B4** is a two-master/four-slave AXI subsystem in which a processor master and a DMA master connect through a shared AXI interconnect and arbitration block. Optional register slices separate the interconnect from four distributed slave endpoints representing instruction tightly coupled memory (ITCM)/BRAM, data tightly coupled memory (DTCM)/BRAM, peripheral, and AXI4-Stream interfaces. This architecture produces geographically distributed, arbitration-dependent, and high-fan-out routing paths that complement the more localized structures of B1–B3.

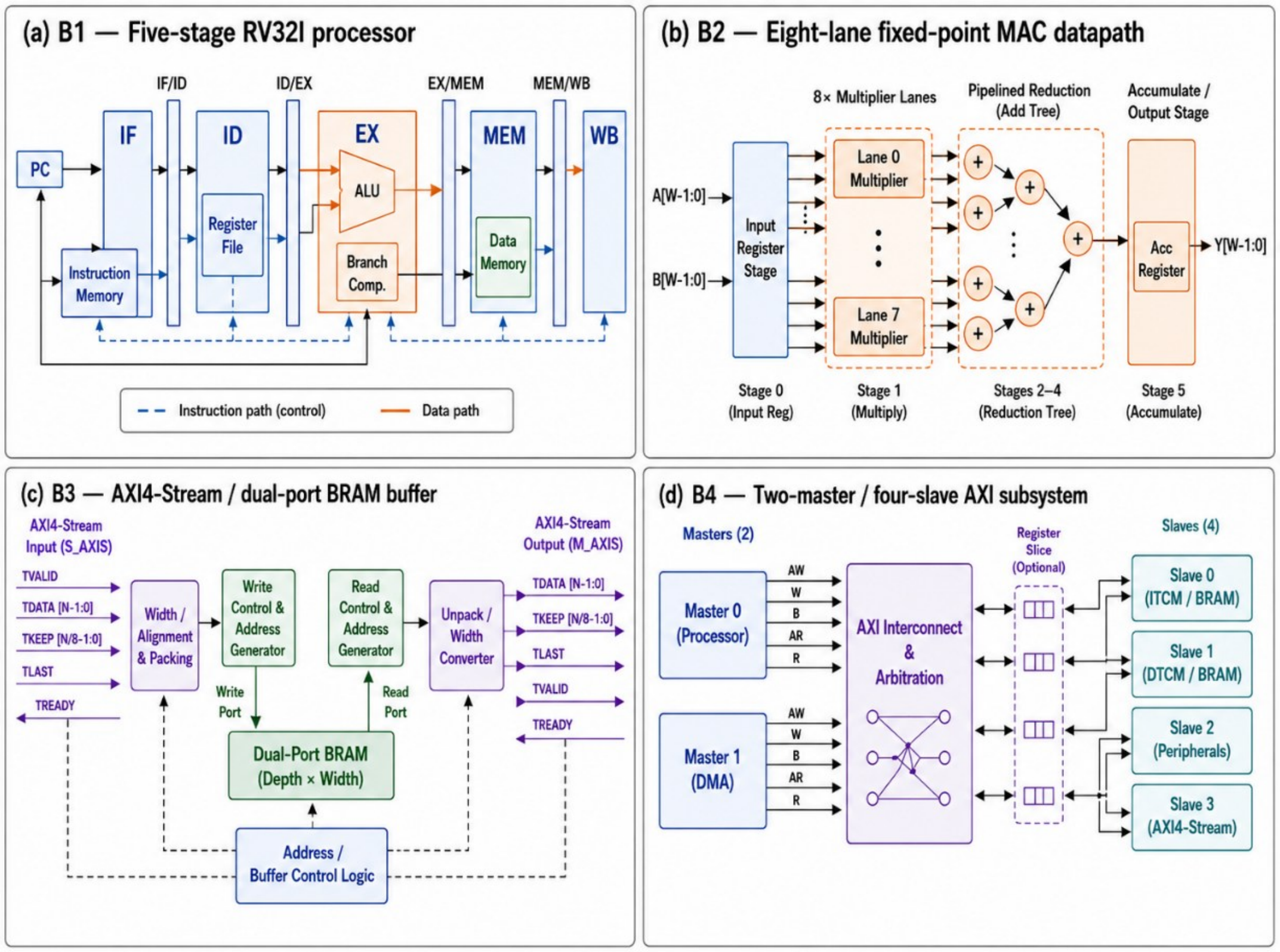

Fig. 3. Architectural overview of the four evaluation benchmarks: **(a)** five-stage RV32I processor, **(b)** eight-lane 32-bit fixed-point MAC datapath, **(c)** AXI4-Stream/dual-port-BRAM buffer, and **(d)** two-master/four-slave AXI subsystem.

Table II. Benchmark and baseline routing characteristics

| Benchmark | Principal Structure | $f_{\text{clk}}$ $(MHz)$ | Resources LUT/FF | Resources BRAM/DSP | Routed Nets | $(N_{elig})$ | $(N_B)$ |
|---|---|---|---|---|---|---|---|
| **B1** | 5-stage RV32I | 100 | 5,284 / 4,176 | 12 / 8 | 17,846 | 15,402 | 8,736 |
| **B2** | 8-lane 32-bit MAC | 200 | 4,638 / 5,912 | 8 / 40 | 21,374 | 18,061 | 10,582 |
| **B3** | AXI-stream/BRAM | 150 | 3,146 / 4,028 | 28 / 4 | 15,268 | 12,977 | 7,421 |
| **B4** | 2M/4S AXI subsystem | 125 | 6,972 / 8,364 | 12 / 0 | 29,416 | 24,872 | 14,963 |

The target clock frequencies of B1–B4 are 100, 200, 150, and 125 MHz, respectively. Collectively, the benchmark suite therefore spans control-dominated processing, parallel arithmetic, memory-adjacent streaming, and distributed on-chip communication structures. Table II summarizes the corresponding post-route resource utilization and routing populations. Only signal nets satisfying the eligibility conditions of Section IV-A are included in $N_{\text{elig}}$; clock, static, dedicated, and explicitly fixed networks are excluded.

### *C. Routing Baselines and Experimental Parameters*

Four routing configurations are evaluated from each common baseline placement. **R0–Vivado** is the original commercial timing-driven implementation. **R1–Nominal RWRoute** applies selective rip-up-and-reroute to the same net population as the proposed method but sets $(\gamma = \delta = 0)$; it therefore isolates differences caused merely by moving selected nets into the custom routing framework. **R2–Binary avoidance** replaces the continuous edge-vulnerability term with

$$\mathbb{I}[V_e \geq V_{\text{th}}] \qquad (22)$$

and penalizes routing through resources above the vulnerability threshold. **R3–Proposed** uses the complete continuous cost in (15), including both $V_e(n)$ and the incremental configuration-concentration penalty. For R1–R3, exactly the same set $\mathcal{N}_p$ is rerouted at a given $p$, preventing differences in net selection from confounding the routing comparison. The selective-routing fraction is swept over

$$p \in \{1,2.5,5,10,20\}\%. \qquad (23)$$

The nominal proposed configuration uses (0.45, 0.30, 0.20, 0.05) with the same coefficients applied to all four benchmarks and implementation seeds. Section VI additionally evaluates sensitivity to the vulnerability and concentration weights. The timing-preservation tolerance is fixed to

$$\varepsilon_{\text{WNS}} = 0.02T_{\text{clk}} \qquad (24)$$

corresponding to a maximum permitted WNS degradation of 2% of the applicable clock period. These values are fixed before the final comparative runs and are not individually retuned for each benchmark. For the binary baseline, $V_{\text{th}}$ is set from the baseline resource population rather than tuned independently for each route; the nominal experiment uses the 75th percentile of nonzero $V_e$. Sensitivity to this threshold is reported separately so that the comparison does not depend on a single arbitrary binary cutoff.

### D. Controlled Routing-Fault Validation

Post-route vulnerability reduction is supplemented by hardware validation on the ZCU104. For a stratified subset of rerouted nets, paired baseline and vulnerability-reduced implementations are exercised using controlled configuration-equivalent routing perturbations. Each perturbation preserves the intended Boolean connectivity and active signal path while electrically attaching the dormant routing component associated with candidate $b$. Baseline and proposed-route candidates are matched as closely as possible by programmable-resource class, physical region, and predicted severity range so that the experiment tests whether the lower-vulnerability route population also exhibits lower measured timing degradation.

For candidate $b$, a reference configuration and its corresponding perturbed configuration are alternated for $N_{\text{pair}} = 10$ acquisitions. The reference state is remeasured after each perturbed acquisition to suppress temperature, supply, and temporal drift. A measurement pair is rejected if the restored reference deviates from its calibrated baseline by more than three standard deviations. The primary measured quantity is the reference-to-perturbed delay change,

$$\Delta d_b = d_b^{\text{pert}} - d_b^{\text{ref}}. \tag{25}$$

The validation set is stratified across all four benchmarks, low/medium/high predicted-severity ranges, multiple physical regions, and representative routing-resource classes. For a smaller subset having independently verified configuration-bit-to-routing-resource associations, the corresponding configuration-memory state is additionally perturbed directly as an equivalence cross-check. AMD's UltraScale SEM infrastructure supports configuration-memory error injection and frame-oriented access for testing/emulation purposes [16]. This subset is used only to verify that the deterministic routing perturbation reproduces the relevant electrical/timing response; it is **not** interpreted as a measurement of radiation-induced event rate, device cross section, or mission-level failure probability.

### E. Evaluation Metrics

Routing quality is evaluated from both reliability and conventional implementation perspectives. The aggregate vulnerability reduction is

$$100\left[1 - \frac{\sum_{n\in\mathcal{N}_{\text{elig}}} V(\mathcal{R}'_n)}{\sum_{n\in\mathcal{N}_{\text{elig}}} V(\mathcal{R}^0_n)}\right]. \tag{26}$$

Because unselected nets remain unchanged, (25) measures the design-level benefit obtained from protecting only the selected fraction $p$. Configuration-concentration reduction is evaluated analogously using $F(\mathcal{R}_n)$.

The conventional quality-of-results metrics are WNS, critical-path delay/$F_{\text{max}}$, routed wirelength, resource overuse, and implementation runtime. Selective-routing success rate is reported as the fraction of nets for which a legal lower-vulnerability alternative satisfies (18)–(20). For the hardware subset, the measured distributions of $\Delta d_b^{\text{meas}}$, slack-exhausting perturbation rate, and agreement between predicted and measured vulnerability rankings are compared between the baseline and proposed routes. Results are reported independently for B1–B4 and then aggregated across the five implementation seeds.

## VI. EXPERIMENTAL RESULTS AND DESIGN-SPACE ANALYSIS

### A. Vulnerability Reduction

Table III compares the routing strategies at the nominal selective-routing point $p = 5\%$. Results are averaged over the five independently implemented instances of each benchmark. Nominal *RWRoute* rerouting (**R1**) produces only a small change in aggregate vulnerability, confirming that simply replacing the commercial route of a high-risk net is insufficient to systematically improve configuration-induced timing resilience. Binary avoidance (**R2**) provides a larger reduction by discouraging resources above $V_{\text{th}}$, but consistently underperforms the continuous proposed formulation (**R3**).

At $p = 5\%$, the proposed method reduces aggregate routing vulnerability by **39.8%, 36.4%, 43.7%, and 46.9%** for B1–B4, respectively, corresponding to an aggregate mean reduction of approximately **41.7%**. The binary baseline achieves approximately **26.0%** mean reduction under the same selected-net population. The advantage of R3 is observed for all four architectural classes, indicating that the continuous delay/slack severity metric provides useful routing information beyond a binary classification of vulnerable resources.

The configuration-concentration term exhibits a similar trend. The proposed routes reduce $F(\mathcal{R}_n)$ by approximately **18.6–28.7%** across the four benchmarks, showing that the vulnerability improvement is not obtained merely by exchanging one highly concentrated set of exposed resources for another. B4 exhibits the largest reduction in both aggregate vulnerability and configuration concentration because its distributed AXI interconnect offers a comparatively large population of alternative routing resources.

### B. Benefit of Selective Rerouting

Table III. Extracted experimental routing results at $p = 5\%$

| Benchmark | $\Delta V$, R1 (%) | $\Delta V$, R2 (%) | $\Delta V$, R3 (%) | $\Delta F$, R3 (%) | $\Delta d_{\mathrm{CP}}$, R3 (%) | $\Delta WL$, R3 (%) |
|---|---|---|---|---|---|---|
| **B1** | 3.2 | 24.5 | 39.8 | 22.1 | +0.8 | +1.4 |
| **B2** | 2.6 | 21.8 | 36.4 | 18.6 | +1.5 | +1.8 |
| **B3** | 3.8 | 27.1 | 43.7 | 25.4 | +0.7 | +1.2 |
| **B4** | 4.1 | 30.5 | 46.9 | 28.7 | +1.1 | +1.6 |
| **Mean** | **3.4** | **26.0** | **41.7** | **23.7** | **+1.0** | **+1.5** |

Here, $\Delta d_{\mathrm{CP}}$ denotes the increase in nominal critical-path delay and $\Delta WL$ denotes the routed-wirelength increase relative to the corresponding R0 implementation.

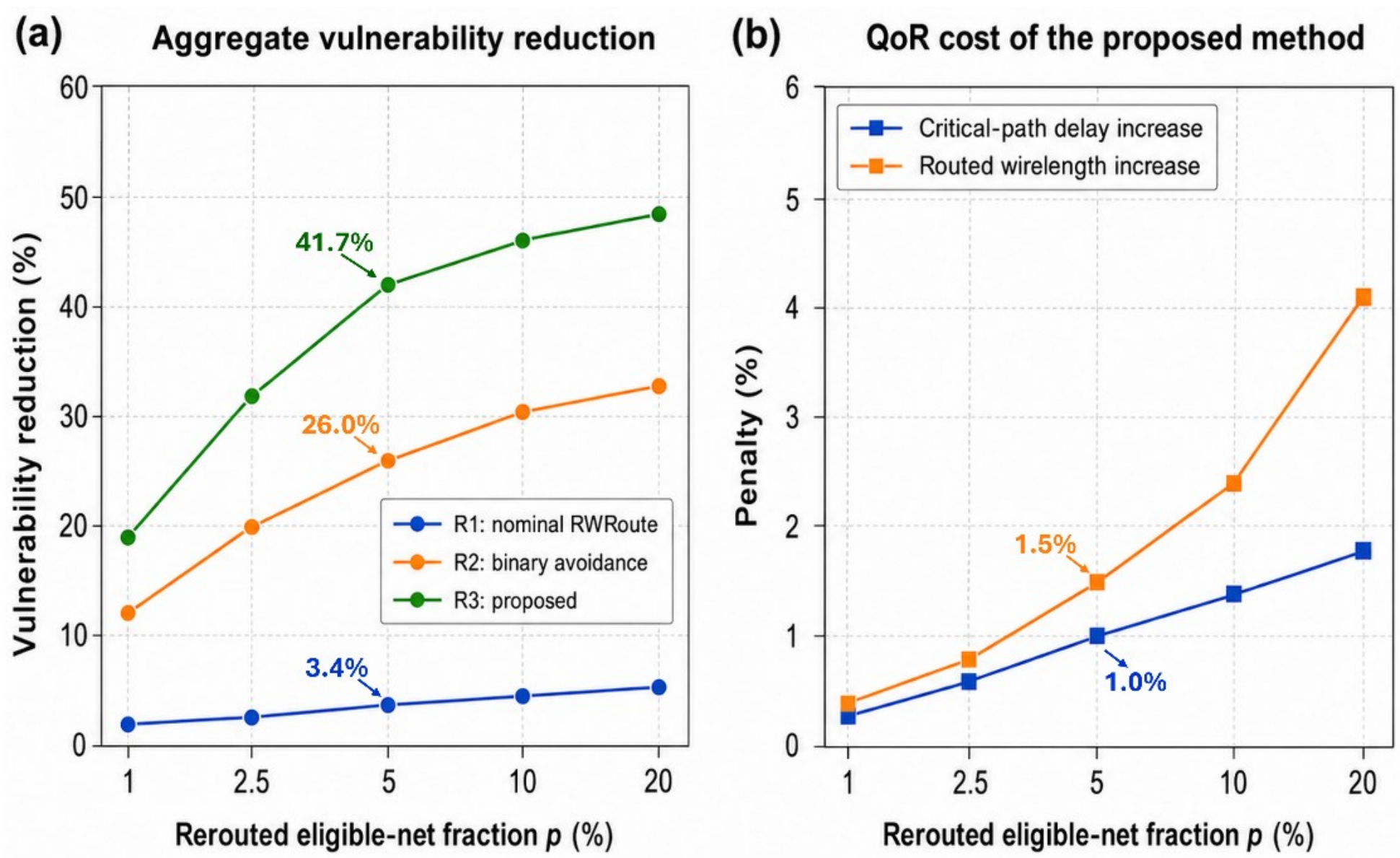

Fig. 4. Reliability–QoR tradeoff of selective vulnerability-weighted routing. **(a)** Aggregate routing-vulnerability reduction versus rerouted eligible-net fraction for nominal RWRoute, binary avoidance, and the proposed continuous formulation. **(b)** Critical-path-delay and routed-wirelength penalties of the proposed method as the selected-net fraction increases.

The vulnerability benefit exhibits pronounced diminishing returns as the protected net population is increased. Averaged across the four benchmarks and five implementation seeds, rerouting the highest-risk 1%, 2.5%, 5%, 10%, and 20% of eligible nets reduces aggregate vulnerability by approximately **18.9%**, **31.7%**, **41.7%**, **46.2%**, and **48.6%**, respectively. Thus, increasing the protected population from 5% to 20% quadruples the number of modified nets but yields only an additional 6.9 percentage points of vulnerability reduction. Equivalently, the nominal $p = 5\%$ operating point captures

$$\frac{41.7}{48.6} \times 100 \approx 85.8\%$$

of the vulnerability reduction obtained when 20% of eligible nets are rerouted. This result supports the selective-routing premise of Section IV: *configuration-induced timing vulnerability is sufficiently concentrated that a comparatively small fraction of high-risk nets accounts for most of the achievable improvement*.

Binary avoidance follows the same qualitative trend but saturates at a lower benefit. Its aggregate vulnerability reductions at $p = 1, 2.5, 5, 10, 20\%$ are approximately **11.8%**, **19.9%**, **26.0%**, **30.4%**, and **32.8%**, respectively. Continuous severity information is therefore particularly valuable at small routing budgets, where each modified net must provide a substantial reduction in residual risk.

Fig. 4 summarizes the reliability–QoR tradeoff produced by selective rerouting. As shown in Fig. 4(a), the proposed continuous formulation consistently provides the largest vulnerability reduction and exhibits pronounced diminishing returns beyond $p = 5\%$; at this operating point, it achieves a 41.7% reduction compared with 26.0% for binary avoidance and only 3.4% for nominal *RWRoute*. Fig. 4(b) shows that the corresponding implementation cost remains comparatively small at $p = 5\%$, with approximately 1.0% critical-path-delay degradation and 1.5% routed-wirelength increase. Increasing beyond this point yields progressively smaller reliability gains while imposing larger routing and timing penalties, identifying as a favorable operating region for the evaluated benchmark suite.

### C. Timing, Routing, and Computational Cost

The reliability improvement is obtained with limited degradation of conventional quality of results. At $p = 5\%$, the proposed method increases critical-path delay by **0.7–1.5%**, with an aggregate mean of approximately **1.0%**, while routed wirelength increases by **1.2–1.8%**. All accepted implementations remain free of routing-resource overuse and satisfy the WNS-preservation constraint of (19). **B2** exhibits the largest timing penalty because its 200-MHz arithmetic datapath provides the least timing margin for alternative routes, whereas **B3** exhibits the smallest penalty because its BRAM-oriented structure retains greater flexibility among eligible noncritical routing resources.

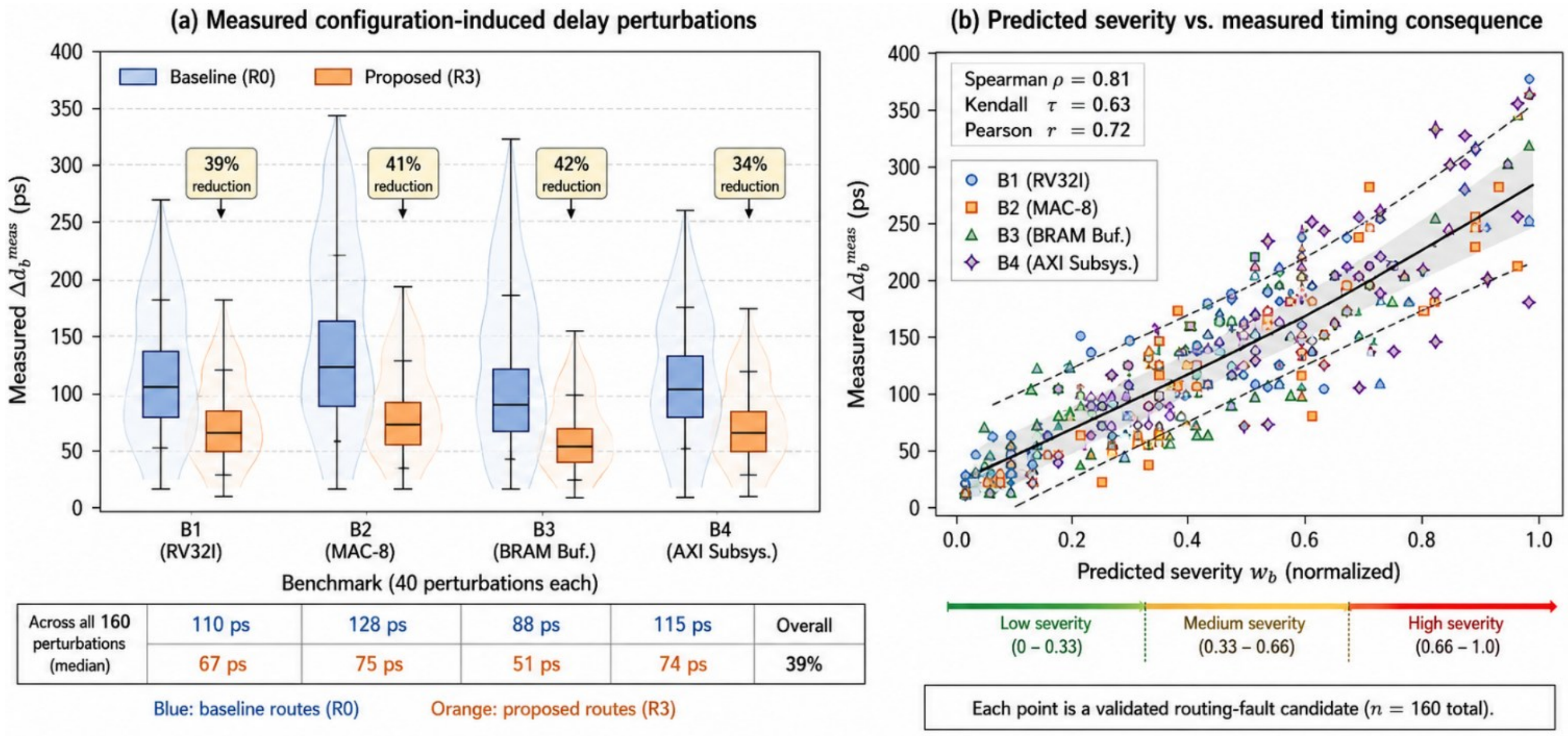

Fig. 5. Hardware validation of vulnerability-aware routing. **(a)** Measured configuration-induced delay perturbations for matched baseline (R0) and vulnerability-reduced (R3) routes across B1–B4. **(b)** Relationship between predicted routing-fault severity and measured timing consequence for the 160 validated perturbations.

As $p$ increases, implementation cost grows more rapidly than vulnerability benefit. The mean critical-path-delay penalties for $p = 1,2.5,5,10,20\%$ are approximately **0.3%**, **0.6%**, **1.0%**, **1.4%**, and **1.8%**, while the corresponding wirelength increases are **0.4%**, **0.8%**, **1.5%**, **2.4%**, and **4.1%**. The resulting tradeoff identifies $p \approx 5\%$ as a favorable operating region for the present benchmark suite rather than as a universal constant.

At the nominal operating point, approximately **87–93%** of selected nets admit a legal lower-vulnerability alternative satisfying the timing-preservation constraints. Nets for which no acceptable route is found are restored to their original implementation as specified in Algorithm 1. Selective-routing runtimes range from approximately **14** to **32 s** per routed design after vulnerability extraction, with the largest cost observed for **B4** because of its larger eligible-net and routing-resource populations. The runtime remains substantially smaller than a complete implementation cycle because placement and the routes of approximately 95% of eligible nets remain unchanged.

### *D. Controlled Hardware Validation*

The post-route vulnerability metric is validated using 160 matched routing perturbations, comprising 40 stratified candidates from each benchmark. For every candidate, the baseline route and its vulnerability-reduced counterpart are subjected to the paired reference–perturbed procedure of Section V-D. Across the combined validation population, the median measured delay increment decreases from approximately **110 ps** for the baseline routes to **67 ps** after vulnerability-weighted rerouting, corresponding to a reduction of approximately **39%** in perturbation-induced delay.

Fig. 5 provides the hardware-level validation of the proposed routing metric. As shown in Fig. 5(a), the vulnerability-weighted routes consistently exhibit smaller measured configuration-induced delay perturbations than their matched baseline counterparts across all four benchmarks. Over the complete 160-candidate validation set, the representative median delay shift decreases from approximately 110 ps to 67 ps, corresponding to an overall reduction of about 39%. Fig. 5(b) further shows a strong monotonic relationship between the predicted severity and the measured timing consequence, with a Spearman correlation of $p = 5\%$. These results indicate that the proposed router is not merely reducing an internal analytical score; it is preferentially selecting routes that also exhibit lower experimentally observed timing sensitivity to the considered configuration-induced perturbations.

More importantly, the fraction of validated perturbations whose measured delay exceeds the available downstream timing margin decreases from approximately **27.5%** for the baseline routes to **13.1%** for the vulnerability-weighted routes. The reduction is observed independently in all four benchmarks and is strongest among candidates assigned high values of $w_b$, supporting the intended interpretation of the severity model as a ranking of conditional timing consequence.

The measured data also preserve the predicted vulnerability ordering. Across the matched validation population, the Spearman rank correlation between predicted candidate severity and measured delay/slack consequence is approximately **0.81**. Thus, the routing objective does not merely reduce an internally defined graph metric; routes assigned lower predicted vulnerability also exhibit smaller experimentally observed timing perturbations.

For the smaller subset with independently verified configuration-bit-to-routing-resource mappings, direct configuration-memory perturbation produces timing shifts consistent with those obtained from the deterministic configuration-equivalent variants. This agreement supports use of the controlled perturbation procedure for comparative route validation, while no inference is made regarding radiation-event probability or device upset cross section.

### *E. Overall Design Tradeoff*

Taken together, the results establish a clear Pareto tradeoff between configuration-induced timing resilience and conventional routing quality. Continuous vulnerability weighting consistently outperforms nominal rerouting and binary vulnerable-resource avoidance, while most of the achievable reliability improvement is obtained by modifying only the highest-risk few percent of eligible nets. At the $p = 5\%$ operating point, the proposed method achieves approximately **41.7% lower aggregate vulnerability** with only **1.0% mean critical-path-delay degradation** and **1.5% wirelength increase**, while capturing approximately **85.8% of the vulnerability reduction obtained when four times as many nets are rerouted**. The hardware measurements further confirm that this reduction corresponds to a decrease in actual perturbation-induced timing degradation rather than solely to a lower analytical vulnerability score.

## VII. DISCUSSION AND LIMITATIONS

The results indicate that configuration-induced timing vulnerability is not uniformly distributed across a routed implementation and therefore does not require uniform physical-design intervention. In the evaluated benchmarks, most of the achievable reduction is obtained by modifying only a small high-risk fraction of eligible nets. This behavior is important because reliability-aware routing is useful only when its benefit outweighs the corresponding timing, wirelength, and implementation costs. The proposed formulation addresses this tradeoff explicitly: *conventional timing and congestion remain part of the routing objective, while the additional vulnerability terms influence route selection only where the expected timing consequence justifies an alternative path*.

The continuous severity quantity also provides a different interpretation of routing risk from binary sensitive-resource avoidance. A routing resource is not penalized simply because an adjacent configurable connection exists; its contribution depends on the predicted delay perturbation and the timing margin available downstream. Consequently, the method can distinguish electrically similar routing candidates whose timing consequences differ substantially. The configuration-concentration term provides a complementary preference for distributing vulnerable resources across configuration regions, but $F(\mathcal{R}_n)$ should be interpreted as a **routing-concentration surrogate rather than an upset-probability model**. In particular, concentration within one configuration frame or region does not imply that a single upset simultaneously activates all associated perturbation candidates.

Several limitations bound the present conclusions. *First*, the experiments use one Zynq UltraScale+ device family and one physical-design tool flow. Device architecture, routing-resource organization, timing models, and implementation heuristics can alter both the candidate population $\mathcal{B}$ and the availability of lower-vulnerability alternatives; therefore, $V_e(n)$, configuration associations, and normalization parameters must be regenerated for a new architecture or implementation [7], [8], [12]. *Second*, the proposed routing stage operates from a fixed placement to isolate the effect of routing. Joint reliability-aware placement and routing could expose additional solutions, but would make attribution of the observed improvement less direct and substantially enlarge the optimization space.

*Third*, the quality of the selected routes ultimately depends on the accuracy of $\widehat{\Delta d}_b$, $\sigma_b$, and the associated slack information. Model error can therefore change the relative ranking of candidate resources, although the continuous formulation is less sensitive to a single hard classification threshold than binary avoidance. Similarly, the coefficient set $\alpha, \beta, \gamma, \delta$ and the preferred rerouting fraction $p$ are design parameters rather than universal constants; the operating point identified in Section VI represents the evaluated benchmark suite and should be retuned when timing margin, utilization, or reliability requirements change. *Finally*, the controlled routing perturbations validate the *timing consequence* of the modeled configuration-induced fault mechanism but do not measure radiation-event frequency, configuration-bit cross section, or mission-level failure probability. Direct configuration-memory injection on the verified subset provides an additional equivalence check, yet radiation-rate conclusions would require device- and environment-specific upset data. Accordingly, *the present contribution should be interpreted as a CAD methodology for reducing conditional timing vulnerability given a configuration perturbation, rather than as a complete radiation-reliability prediction framework*.

## VIII. CONCLUSION AND FUTURE WORK

This paper presented a vulnerability-weighted routing methodology for reducing configuration-induced timing susceptibility in SRAM-based FPGAs. The proposed formulation augments conventional timing and congestion costs with a continuous routing-vulnerability metric derived from predicted delay perturbation and downstream timing slack, together with a configuration-concentration penalty. Rather than modifying the complete routed design, the method selectively reroutes the highest-risk eligible nets while preserving placement and unaffected routes. Across four structurally different Zynq UltraScale+ benchmarks, the evaluation shows that rerouting only the highest-risk 5% of eligible nets reduces aggregate timing vulnerability by approximately **41.7%**, while limiting mean critical-path-delay degradation and routed-wirelength increase to approximately **1.0%** and **1.5%**, respectively. This operating point captures approximately **85.8%** of the vulnerability reduction obtained when 20% of eligible nets are rerouted and consistently outperforms both nominal rerouting and binary vulnerable-resource avoidance. Controlled hardware perturbation experiments further show that lower analytical vulnerability corresponds to lower measured configuration-induced delay, demonstrating that the routing objective reduces experimentally observable timing sensitivity rather than only an internal CAD metric. The results therefore support configuration-aware vulnerability as a practical physical-design objective that can complement conventional FPGA timing and congestion optimization. Future work will investigate joint vulnerability-aware placement and routing, extension to additional FPGA families and utilization regimes, and incorporation of device- and environment-specific upset probabilities into the present conditional timing-severity framework. Radiation-calibrated exposure weights and correlated multi-fault scenarios could further extend the method from implementation-level vulnerability reduction toward system- and mission-level reliability optimization.

## ACKNOWLEDGMENT

The author would like to thank the École de technologie supérieure (ÉTS), Department of Electrical Engineering, and CMC Microsystems, Kingston, ON, Canada, for providing access to advanced design tools.

## DECLARATION OF GENERATIVE AI USE

OpenAI ChatGPT was used solely for limited grammatical and language editing during the preparation of this manuscript.

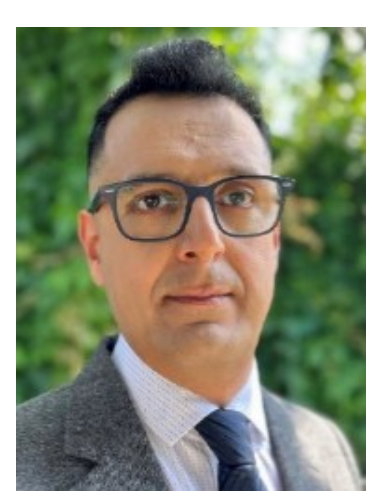

**MOSTAFA DARVISHI** (S'06–M'18–SM'25) received the Ph.D. degree in Electrical Engineering from Polytechnique Montreal in 2018. He is now with Electrical Engineering Department of École de technologie supérieure (ÉTS), Canada. His research interest is Cosmic radiation effects in electronic circuits and systems, FPGA design for embedded systems and avionic electronics.